\documentclass[letterpaper]{article} 
\usepackage{aaai21}  
\usepackage{times}  
\usepackage{helvet} 
\usepackage{courier}  
\usepackage[hyphens]{url}  
\usepackage{graphicx} 
\def\UrlFont{\rm}  
\usepackage{natbib}  
\usepackage{caption} 
\usepackage{graphicx}
\usepackage{comment}
\usepackage{amsmath,amssymb} 
\usepackage{color}
\usepackage{epsfig}
\usepackage{stmaryrd}
\usepackage{algorithm}
\usepackage{algorithmic}
\usepackage[breaklinks=true,bookmarks=false]{hyperref}

\newcommand{\vect}[1]{\mbox{\boldmath$#1$}}
\title{Efficient CNN Building Blocks for Encrypted Data}
\author{
        Nayna Jain\textsuperscript{\rm 1,4},
        Karthik Nandakumar\textsuperscript{\rm 2},
        Nalini Ratha\textsuperscript{\rm 3},
        Sharath Pankanti\textsuperscript{\rm 5},
        Uttam Kumar \textsuperscript{\rm 1}\\}
\affiliations{
    \textsuperscript{\rm 1} Center for Data Sciences, International Institute of Information Technology, Bangalore  \\
    \textsuperscript{\rm 2} Mohamed Bin Zayed University of Artificial Intelligence \\
    \textsuperscript{\rm 3} University at Buffalo, SUNY\\
    \textsuperscript{\rm 4} IBM Systems \\
    \textsuperscript{\rm 5} Microsoft \\
    karthik.nandakumar@mbzuai.ac.ae,
    nratha@buffalo.edu, \\ nayna.jain@iiitb.org/nayna.jain@ibm.com,
    sharath.pankanti@gmail.com, uttam@iiitb.ac.in}

\begin{document}



\maketitle

\begin{abstract}
Machine learning on encrypted data can address the concerns related to privacy and legality of sharing sensitive data with untrustworthy service providers, while leveraging their resources to facilitate extraction of valuable insights from otherwise non-shareable data. Fully Homomorphic Encryption (FHE) is a promising technique to enable machine learning and inferencing while providing strict guarantees against information leakage. Since deep convolutional neural networks (CNNs) have become the machine learning tool of choice in several applications, several attempts have been made to harness CNNs to extract insights from encrypted data. However, existing works focus only on ensuring data security and ignore security of model parameters. They also report high level implementations without providing rigorous analysis of the accuracy, security, and speed trade-offs involved in the FHE implementation of generic primitive operators of a CNN such as convolution, non-linear activation, and pooling. In this work, we consider a Machine Learning as a Service (MLaaS) scenario where both input data and model parameters are secured using FHE. Using the CKKS scheme available in the open-source HElib library, we show that operational parameters of the chosen FHE scheme such as the degree of the cyclotomic polynomial, depth limitations of the underlying leveled HE scheme, and the computational precision parameters have a major impact on the design of the machine learning model (especially, the choice of the activation function and pooling method). Our empirical study shows that choice of aforementioned design parameters result in significant trade-offs between accuracy, security level, and computational time. Encrypted inference experiments on the MNIST dataset indicate that other design choices such as ciphertext packing strategy and parallelization using multithreading are also critical in determining the throughput and latency of the inference process.
\end{abstract}

\section{Introduction}

Deep neural networks have proven to be a promising technology due to their ability to achieve competitive machine learning performance in diverse domains including computer vision. Ever since AlexNet achieved a top-5 error rate of 16.4\% using Convolutional Neural Networks (CNN) for the image classification task on the ImageNet dataset of 1.2 million high-resolution images as part of ILVSRC 2012 \cite{NIPS2012_4824} competition, various CNN architectures \cite{vggnet}\cite{googlenet}\cite{zfnet} have been developed to improve accuracy and other performance metrics. Though the continuous improvement in the performance of CNNs has hastened their adoption in broader computer vision applications such as mobile and embedded devices \cite{DBLP:journals/corr/HowardZCKWWAA17}\cite{Phan_2020_WACV}, the large computing/memory resource requirements of CNNs present practical challenges in applying them in many applications. This is often overcome by outsourcing the network training and/or inference computations to the Cloud, which is commonly referred to as Machine Learning as a Service (MLaaS).

\begin{figure}
\begin{center}
\includegraphics[width=0.5\textwidth]{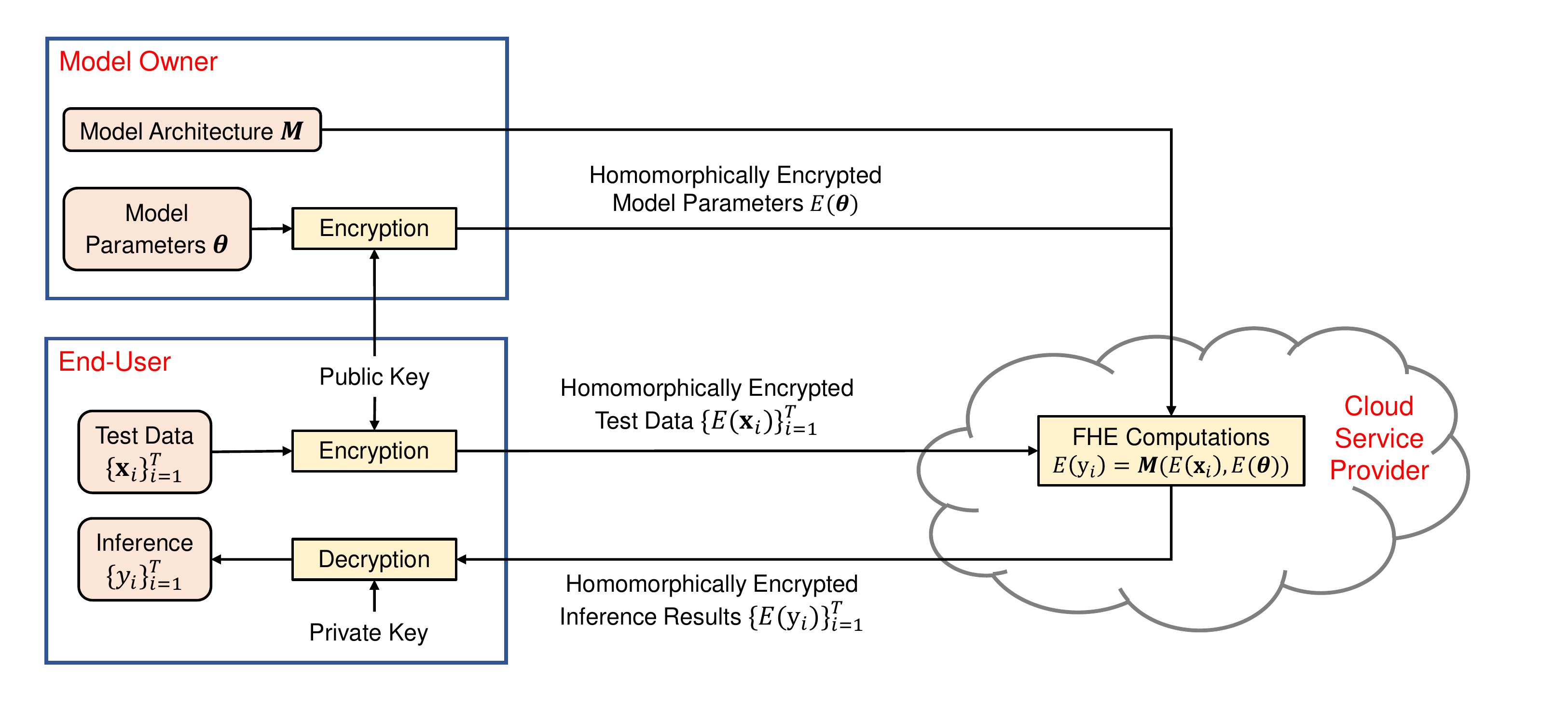}
\end{center}
\caption{In a conventional Machine Learning as a Service (MLaaS) scenario, both the data and model parameters are unencrypted. Existing works on secure inference assume only the data is encrypted. Our proposed approach encrypts both the data and model parameters.}
\label{fig:MLaaS}
\end{figure}

In MLaaS, there are three parties: (i) the end-user, who requires predictions on sensitive data, (ii) the cloud service provider, who has access to large computational resources, and (iii) the model owner, who has developed the inference model, trained using machine learning algorithms that may be proprietary. In conventional MLaaS scenarios, the cloud service providers typically have access to the inference models as well as the content of the query data. In many applications, such access to data and the model is undesirable because the query data/models could reveal sensitive information. Moreover, sharing of such data may be prohibited under emerging privacy regulations such as GDPR.  This has led to the development of a variety of privacy preserving machine learning algorithms. In this paper, we focus on a specific class of privacy preserving machine learning algorithms that rely on Homomorphic Encryption (HE).

Most of the approaches presented in the literature for encrypted inference encrypt only the input data and do not encrypt the model parameters. This is based on the assumption that the same entity owns the model and runs the cloud service. However, access to the model parameters could lead to competitive disadvantage for the model owner, when the model owner and cloud service provider are separate entities (e.g., a start-up providing the MLaaS service on a Amazon or Google cloud). Hence, it becomes imperative to protect both the input data and model parameters in such scenarios (see Figure \ref{fig:MLaaS}). \cite{encryptedmodelinproceedings} discussed about encrypted models in the context of Hyperplane Decision, Navie Bayes and Decision Tree algorithm. In this work, we focus on the scenario where the model owner and cloud service provider are separate entities and the cloud provider is not trusted. Therefore, we encrypt the model parameters\footnote{Note that the model architecture is still available to the cloud service provider in the clear.} in addition to the input data using the end-user's public key (as shown on Figure \ref{fig:MLaaS}). The advantage of this approach is that the cloud service provider cannot derive any benefit from the model directly. The disadvantage of model parameter re-encryption using the specific client's public key is the increased computation and communication burden of the model owner.

\subsection{Contributions}
\label{subsec:Contributions}

The objective of this work is to develop the component building blocks for enabling generic CNN inference on encrypted data. We use the implementation of the CKKS scheme available in the HElib library \cite{helib} for all our analysis. However, we note that the lack of a bootstrapping implementation for CKKS makes it impossible to perform inferencing on any arbitrary CNN. Hence, we design depth-constrained CNNs that operate within the multiplicative depth limitations of the underlying leveled HE scheme. Our contributions are four-fold:

\begin{enumerate}
\item MLaaS scenario for convolution neural networks is proposed where both the input data and model parameters are encrypted, thereby enabling a model owner to provide the MLaaS service on an untrusted cloud.

\item CNN inference on encrypted data using the CKKS scheme is presented, which eliminates the need for careful model parameter quantization.

\item Implementation and performance analysis of illustrative linear operators such as convolution and matrix multiplication as well as non-linear operations such as ReLU and maxpooling in the encrypted domain is provided.

\item Various multithreading strategies were explored and the analysis revealed that the optimal strategy involves a complex interplay between system resources and the computations involved.

\end{enumerate}

\section{Background and Related Work}

\subsection{Homomorphic Encryption}
Homomorphic encryption schemes enable a class of computations on the ciphertexts (i.e., encrypted data) without decrypting them \cite{HESurvey}. Let $\llbracket x \rrbracket$ denote the encryption of value $x$ using a public key $pk$. A cryptosystem is said to be \emph{fully homomorphic} (FHE) \cite{gentry2009fully} if it enables computation of any arbitrary function $f(x_1,x_2,\cdots,x_d)$ in the encrypted domain. In other words, there exists a function $g$ such that $f(x_1,x_2,\cdots,x_d) = \mathcal{D}\left(g(\llbracket x_1 \rrbracket, \llbracket x_2 \rrbracket,\cdots \llbracket x_d \rrbracket), sk\right)$, where $\mathcal{D}$ represents the corresponding decryption mechanism of the cryptosystem using private/secret key $sk$.

Recently, a number of FHE schemes have been proposed based on the hardness of the Ring Learning With Errors (RWLE) problem. The most well-known examples of such schemes are BFV \cite{brakerski2012fully}\cite{FV_FHE}, BGV \cite{brakerski2014leveled}, TFHE \cite{chillotti2020tfhe}, and CKKS \cite{cheon2017homomorphic} cryptosystems, all of which support both additive and multiplicative homomorphism. Among the well-known FHE schemes, only CKKS natively supports floating-point operations, albeit with approximate (limited precision) computations. Most of the previous attempts at encrypted inference coarsely quantize the weights of a neural network in order to satisfy the requirements of the underlying HE schemes \cite{gilad2016cryptonets,lou2019she}. In this paper, we explore the use of the CKKS scheme, which eliminates the need for transformations of operands as integers.

FHE is typically obtained by combining a leveled HE scheme with a bootstrapping operation. Since each HE computation increases the ``noise'' level of a ciphertext, a leveled HE scheme has a (parameter-dependent) limit on the computational (multiplicative) depth of the arithmetic circuit that can be computed, while still allowing for meaningful decryption. This limitation is overcome by using Gentry’s bootstrapping technique \cite{gentry2009fully}, which ``refreshes'' the ciphertext and reduces its noise level (at the cost of relying on circular security). While bootstrapping can enable computation of arithmetic circuits with arbitrary multiplicative depth, it comes at a prohibitively high computational cost. For some FHE constructs such as CKKS, there is no existing implementation of true ``Gentry-style'' bootstrapping. This leads to an intractable conundrum. While modern machine (deep) learning algorithms rely on their depth to achieve good generalization performance \cite{poggio2017}, deeper networks are inherently more difficult to compute in the encrypted domain without bootstrapping. Therefore, optimizing the computational depth to achieve an acceptable trade-off between accuracy and practical computational feasibility is one of the fundamental challenges in encrypted deep learning.

\subsection{Secure Machine Learning Using FHE}
\label{sec:related}
Every supervised machine learning algorithm involves two steps: (i) learning the model from training data; (ii) inference on new unknown samples. We presume that an efficient model has already been trained from data available in clear. Our focus is on how to carry out encrypted inferencing with minimal leakage of information for all the parties in MLaaS while providing guaranteed privacy.

Privacy preserving machine learning can have different challenges and threats based on what is required to be private. In \cite{ppmlthreats}, various types of attacks (\emph{e.g.}, reconstruction attacks, model inversion attacks, membership inference attacks) that could potentially reveal private/proprietary information from data/model have been discussed. The work also presents various cryptographic and non-cryptographic approaches to thwart these attacks. Key techniques that have been explored to achieve privacy preserving machine learning include differential privacy, secure multiparty computation (SMPC), and homomorphic encryption. Homomorphic encryption has been considered more suitable for the cloud-based applications, but its practical feasibility is still in question. SMPC-based methods have been generally deemed more practical, but it requires interaction between the client and server resulting in large communication cost. Differential privacy techniques lack the formal guarantees of security under assumptions of more conservative attack models.

Most of the recent work in privacy-preserving machine learning \cite{gilad2016cryptonets}\cite{juvekar2018gazelle}
\cite{fastercryptonets}\cite{dathathri2019chet} has primarily focused on
secure inference where the model has already been trained in the clear.
Once the model has been trained, these methods use either SMPC or homomorphic encryption techniques to enable inference. While CryptoNets \cite{gilad2016cryptonets} rely only on homomorphic encryption to enable inferencing over private input data, Gazelle \cite{juvekar2018gazelle} uses a combination of homomorphic encryption and garbled circuits to achieve  two orders of magnitude speed improvement. Faster CryptoNets \cite{fastercryptonets} have demonstrated improved performance over that of  CryptoNets by leveraging sparsity properties and reduced the amortized inference time from minutes to seconds. CHET \cite{dathathri2019chet} provides an optimizing compiler for Fully Homomorphic Neural Network Inference. It basically provides high level user framework to automate the tuning of parameters for security as well as performance without degrading the accuracy. The framework provides high level and low level intermediate representations that can be linked with different homomorphic libraries. Their results showed that  automated homomorphic circuits from the compiler outperform the circuits that were manually tuned. An encrypted inference service built on top of TensorFlow using secure multiparty computation is discussed in \cite{ppmltensorflow}. Encrypted model was considered in \cite{encryptedmodelinproceedings} for Hyperplane Decision, Naive Bayes, Decision Tree algorithms, but not convolution neural network.

\subsection {Convolutional Neural Network (CNN)}

A CNN is a multi-layered neural network that is usually applied to image/video data. Let $\vect{X}(\ell)$ and $\vect{Z}(\ell)$ denote the input and output of layer $\ell$  ($\ell = 1, 2, \cdots,D$) in the network. Each layer in the network performs specific arithmetic operations on the outputs of the previous layer and passes the result to the subsequent layer, i.e., $\vect{X}(\ell) = \vect{Z}(\ell-1)$ and $\vect{Z}(0)$ is the raw input data. This structure allows higher-level abstract features to be computed as non-linear functions of lower-level features (starting with the raw data). The first few layers of a CNN generally consists of the three types of operations.

\begin{enumerate}
    \item Convolution: This is a linear operation, where the input to the layer\footnote{For the sake of convenience, we drop the layer index $\ell$ in the subsequent discussion.} ($\vect{X}$) is convolved with a kernel ($\vect{W}$) to generate a filtered output $\vect{Y}$. Let $\vect{A}_{i,j}$ denote the element in the $i$-th row and $j$-th column of a matrix $\vect{A}$. Suppose that the size of input $\vect{X}$ is $(M \times N)$ and the size of kernel $\vect{W}$ is $(P \times Q)$. Typically, one can pad the input with zeros to compute the filter responses at the edges. Let $\vect{\widetilde{X}}$ be the zero-padded version of the input $\vect{X}$.

    \begin{equation}
        \label{eqn:conv}
        \vect{Y}_{m,n} = \langle \vect{\widehat{X}}_{m,n},  \vect{\widehat{W}}\rangle,
    \end{equation}

    \noindent where $\langle \vect{a}, \vect{b}\rangle$ is the inner product of two vectors $\vect{a}$ and $\vect{b}$, $\vect{\widehat{W}}$ is the vectorized (flattened) version of the kernel $\vect{W}$, and $\vect{\widehat{X}}_{m,n}$ is the vectorized version of the input window (a window of size $P \times Q$ with $\vect{\widetilde{X}}_{m,n}$ as the top-left element) selected from the padded image. For the sake of simplicity, the stride length is assumed to be $1$ in equation (\ref{eqn:conv}).

    \item Activation: In this step, a point-wise non-linear function is applied to the filter responses. The commonly used activation functions are rectified linear unit (ReLU), sigmoid, and hyperbolic tangent (tanh). In this paper, we consider only the ReLU activation function, which is defined as follows:

     \begin{equation}
        \label{eqn:origReLU}
       ReLU(a) = max(0,a) = \begin{cases} 0, & \mbox{if } a \leq 0 \\ a, & \mbox{if } a > 0. \end{cases}
    \end{equation}

    \item Pooling: This is typically used for dimensionality reduction, where multiple responses within a neighborhood are pooled together. While max pooling is often used for selecting the most dominant response in each neighborhood and is a non-linear operation, mean pooling computes the average response in each neighborhood and is a linear operation. The final output ($\vect{Z}$) of the activation and pooling layers can be modeled as a function $f$ of the filter responses, i.e., $\vect{Z} = f(\vect{Y})$, where $f$ is non-linear.
\end{enumerate}

After the initial layers, the data is flattened into a vector and a few fully connected (FC) layers are added to the network. The output of each node in the FC layer is often computed by applying a non-linear activation function to the weighted average of its inputs, which includes a bias term that always emits value 1. This can be mathematically represented as $\vect{Z} = f(\vect{W}\vect{X})$, where $\vect{W}$ is the weight matrix of the FC layer and $f$ is the activation function. The final layer in a CNN is usually a softmax layer that provides an approximate probability distribution over the class labels.

\section{Proposed Approach}
\label{sec:approach}

In this work, we employ a typical CNN architecture as an illustrative example. Though FHE schemes allow arbitrary computations on the encrypted data, two key challenges need to be circumvented before applying a CNN on encrypted inputs. The first challenge is the appropriate packing of input data to efficiently make use of the Single Instruction Multiple Data (SIMD) operations. The next challenge stems from the inherent inability of most FHE schemes to directly compute non-linear functions (with the exception of polynomials). Often, the non-linear function is approximated by an iterative algorithm or a polynomial function. Since the approximation error is generally inversely proportional to the computational depth (using more iterations or higher-degree polynomials leads to less approximation error), the trade-off between computational depth and accuracy must be carefully managed.

\begin{figure*}[t]
    \begin{center}
    \begin{tabular}{cc}
    \includegraphics[width=0.7\textwidth]{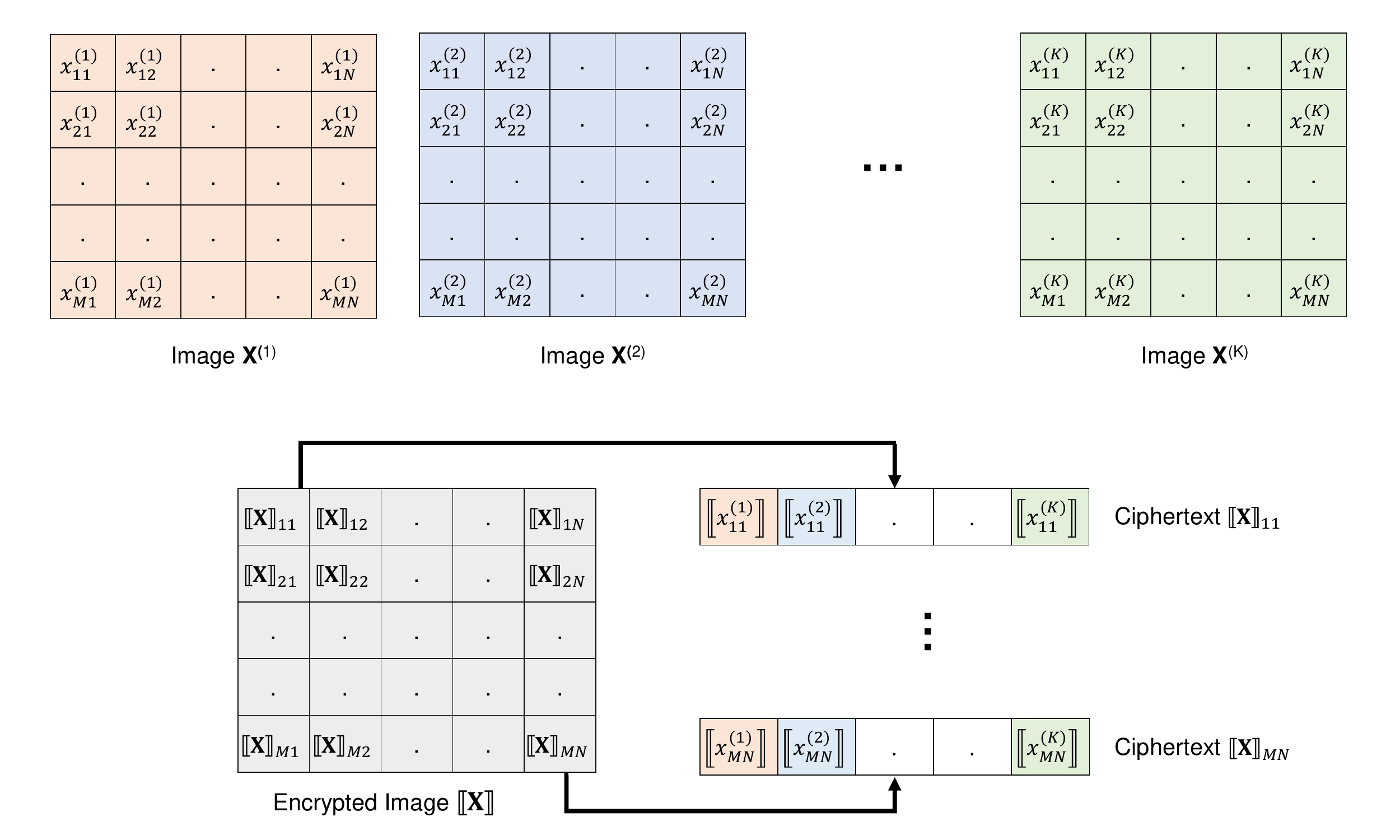} & (a) \\
    \includegraphics[width=0.7\textwidth]{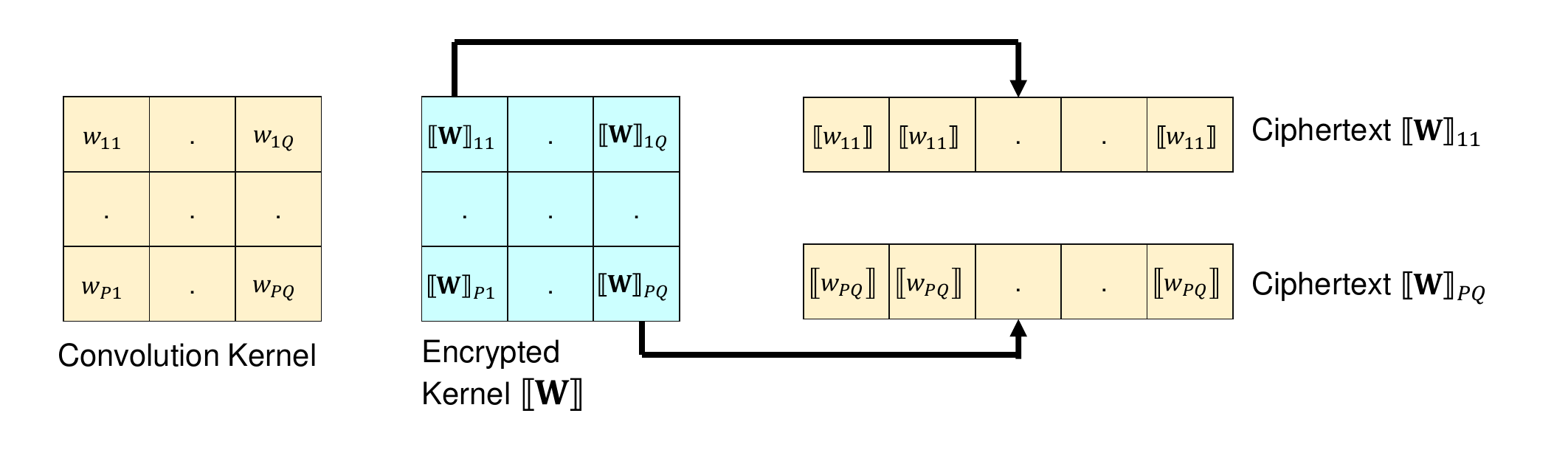} (b) \\
    \end{tabular}
    \caption{Ciphertext packing for batched inference. (a) Ciphertext $\llbracket \vect{X} \rrbracket_{m,n}$ contains the encrypted values of pixel $\llbracket x_{m,n}^{(\cdot)} \rrbracket$ from all the $K$ images in a batch, where $K$ is the number of slots in the ciphertext. (b) Ciphertext $\llbracket \vect{W} \rrbracket_{p,q}$ contains the encrypted value of kernel weight $\llbracket w_{p,q} \rrbracket$ in all the $K$ slots.}
    \label{fig:BatchedInference}
    \end{center}
\end{figure*}

\subsubsection{Convolution Operator:}

We employ the CKKS scheme for encrypted operations. The native plaintext in the CKKS scheme is a polynomial in the cyclotomic ring, which enables packing multiple plaintext values into different ``slots'' in the ciphertext. This ciphertext packing enables parallelization of addition and multiplication operations through SIMD operations. However, it must be emphasized that it is not possible to randomly access values in the individual slots of the ciphertext after packing. Since only a limited set of operations (e.g., rotation of values in the slots) are possible within a ciphertext, the benefits of SIMD operations can be fully realized only when there is minimal interaction between the slots. Therefore, we follow a more conservative and straightforward approach to ciphertext packing that amortizes the computational time by processing a batch of images in parallel.

Let $K$ be the number of slots available in the ciphertext. Given a batch of $K$ images $\vect{X}^{(1)},\vect{X}^{(2)},\cdots,\vect{X}^{(K)}$, where each image is of size $M \times N$, we represent the encrypted image as a ciphertext matrix $\llbracket \vect{X} \rrbracket$ of size $M \times N$ as shown in Figure \ref{fig:BatchedInference}(a). The ciphertext $\llbracket \vect{X} \rrbracket_{m,n}$ contains the encrypted values of pixel $(m,n)$ from all the $K$ images, i.e. $\llbracket \vect{X} \rrbracket_{m,n} = \left[\llbracket x_{m,n}^{(1)} \rrbracket, \llbracket x_{m,n}^{(2)} \rrbracket, \cdots, \llbracket x_{m,n}^{(K)} \rrbracket\right]$. Similarly, the encrypted convolution kernel is also represented as a ciphertext matrix $\llbracket \vect{W} \rrbracket$ of size $P \times Q$, where $P \times Q$ is the kernel size\footnote{Encryption of the convolution kernel addresses the more general scenario where the model provider may be different from the cloud service provider.}. However, as shown in Figure \ref{fig:BatchedInference}(b), the encrypted weight value $\llbracket w_{p,q} \rrbracket$ is repeated across all the slots of the ciphertext $\llbracket \vect{W} \rrbracket_{p,q}$.

While the above simple packing approach greatly increases the memory requirement, it provides great simplicity in terms of implementing the convolution operator. One can directly treat the encrypted image matrix $\llbracket \vect{X} \rrbracket$ and the encrypted kernel $\llbracket \vect{W} \rrbracket$ as equivalent of the plaintext matrices $\vect{X}$ and $\vect{W}$, respectively, and compute the convolution results as the inner product between the vectorized versions of the selected image window and the kernel. Thus, equation (\ref{eqn:conv}) can be modified as follows:

\begin{equation}
    \label{eqn:enc_conv}
        \llbracket \vect{Y} \rrbracket_{m,n} = \langle \llbracket \vect{\widehat{X}} \rrbracket_{m,n},  \llbracket \vect{\widehat{W}} \rrbracket\rangle.
    \end{equation}

When there is no padding involved, the above approach requires $(M-P+1)(N-Q+1)PQ$ ciphertext multiplications (henceforth denoted as CT-CT mult) and ciphertext additions (henceforth denoted as CT-CT add) and consumes a multiplicative depth of $1$.

\subsubsection{Activation and Pooling Operators:}

In this paper, we implement a ReLU function using a polynomial approximation. In \cite{PolyAct}, a number of polynomial approximations have been proposed for the ReLU function. Since one of our objectives is to minimize the computational depth, we choose the following polynomial of degree 2 for our approximation.

\begin{equation} \label{eqn:approxReLU}
    g(u) = 0.47 + 0.50u + 0.09u^2, u \in [-\sqrt{2},\sqrt{2}].
\end{equation}

\noindent A linear transformation may be required to limit $u$ within the appropriate range before the above polynomial approximation is applied. The polynomial approximation of ReLU consumes only a multiplicative depth of 1 and requires 1 CT-CT mult, 2 CT-PT mults (multiplication a ciphertext with a plaintext operand), 2 CT-CT adds, and 1 CT-PT add (addition of a plaintext operand to a ciphertext) operations.

In terms of pooling, we consider mean pooling operator in a $2 \times 2$ neighborhood, which is a linear operation. While mean pooling does not involve any ciphertext multiplication (CT-CT mults), it does require $4$ CT-CT adds and $1$ CT-PT mult. Our activation and pooling operations can be summarized as follows:

\begin{equation}
    \llbracket \vect{Z} \rrbracket_{i,j} = f\left(\llbracket \vect{Y} \rrbracket_{2i-1,2j-1},\llbracket \vect{Y} \rrbracket_{2i,2j-1},\llbracket \vect{Y} \rrbracket_{2i-1,2j},\llbracket \vect{Y} \rrbracket_{2i,2j}\right),
\end{equation}

\noindent where

 \begin{equation}
      \label{eqn:ActPool}
       f(a,b,c,d) = (g(a)+g(b)+g(c)+g(d))/4.
\end{equation}

\section{Experimental Results}
\label{sec:experiments}

\subsection{Environment}
For FHE, we chose IBM's open-source library named HElib \cite{helib} version 1.0.1. Since FHE based experiments require signficant compute capability and efficiency, we chose the natural API interface of HElib library through C++. The configuration details of the computing environment are listed below:

\begin{itemize}
\item System used: Virtual Machine ppc64le POWER9,
\item CPUs: 112 CPUs with 14 sockets, 1 core/socket, 8 threads per core
\item Memory: RAM  - 511 GB,
\item Operating system: Fedora Core 32
\end{itemize}

\subsection{Impact of FHE parameters}
Three key parameters define the operational characteristics of the FHE scheme and determine the security level ($\lambda$). The plaintext space of CKKS is the set of polynomials in the cyclotomic ring $\mathbb{Z}[X]/(\Phi_m(X))$, where $\Phi_m(X)$ is the $m^{th}$ cyclotomic polynomial with degree given by Euler's totient $\phi(m)$. The number of slots ($K$) in the ciphertext is given by $K = \phi(m)/2$. The second key parameter is $L$, which is the bitsize of the modulus of a freshly encrypted ciphertext. Since we are using a leveled HE scheme, $L$ determines the depth of the circuit that can be evaluated without bootstrapping. Increasing $L$ allows more computations (multiplications) to be performed before hitting the noise threshold. Finally, the parameter $r$ determines the computational precision in the encrypted domain. While larger values of $m$ (and hence $\phi(m)$) increase the security level $\lambda$, larger values of $L$ and $r$ generally decrease the security. In our experiments, we choose $m = 2^{16}$ (corresponding to $\phi(m) = 32768$ and $K = 16384$), $r = 35$, and vary $L$ depending on the computational depth required.

\begin{figure*}
\begin{center}
\begin{tabular}{ccc}
    \includegraphics[width=0.3\textwidth]{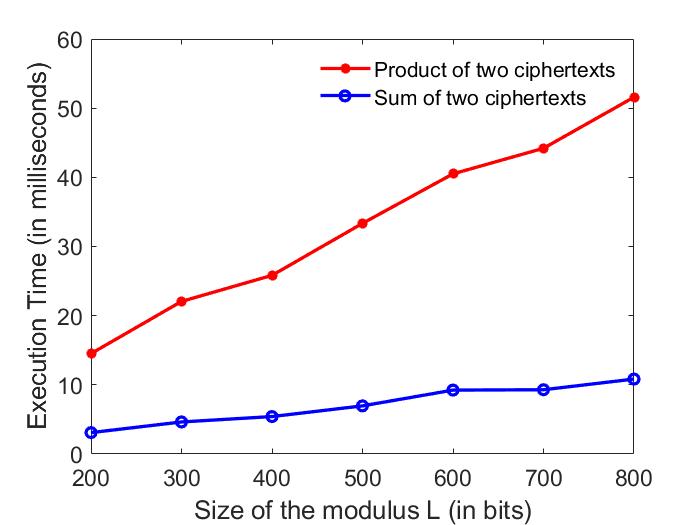}
    & \includegraphics[width=0.3\textwidth]{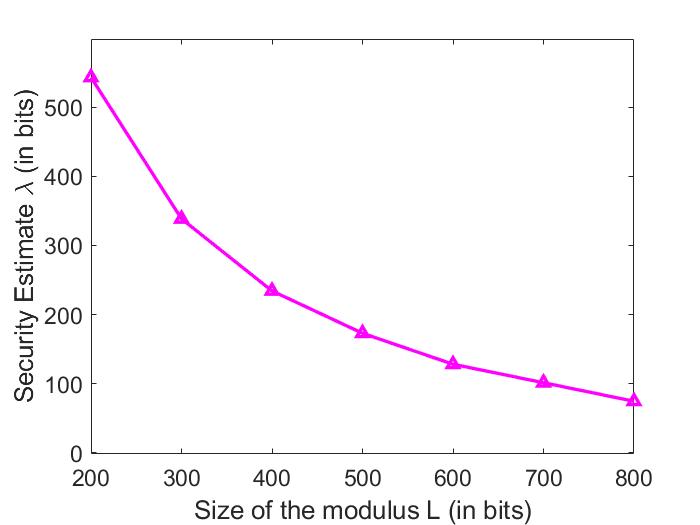}
    & \includegraphics[width=0.3\textwidth]{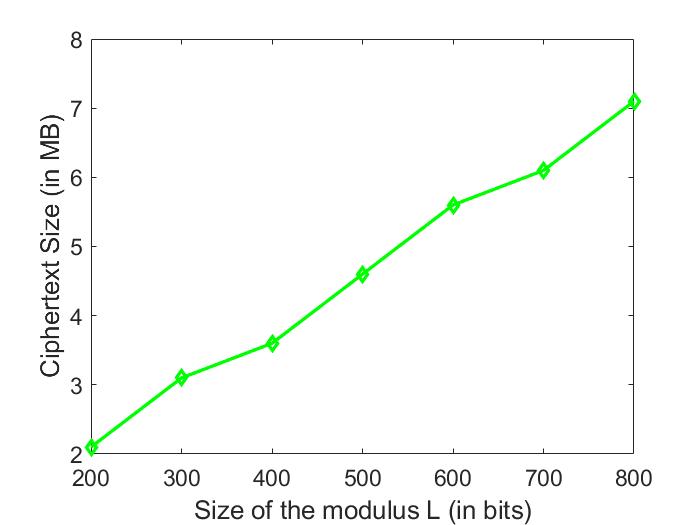}
\\ (a) & (b) & (c)\\
\end{tabular}
\end{center}
\caption{Impact of the depth parameter $L$ on (a) computational time, (b) security level, and (c) ciphertext size. For this experiment, the value of $m$ and $r$ are set to $2^{16}$ and $35$, respectively. The computational time is the average time required for adding/multiplying two ciphertexts over 1000 trials.}
\label{fig:ImpactOfL}
\end{figure*}

We first evaluate the impact of the depth parameter $L$ on the computational time, security level, and ciphertext size. For measuring the computational time, we consider the average time required for the adding/multiplying two ciphertexts (filled with constant values in all the slots) over 1000 trials. As shown in Figure \ref{fig:ImpactOfL}, increasing $L$ increases the computational time and ciphertext size in a linear fashion. Moreover, the rate of increase is much higher for multiplication of two ciphertexts (CT-CT mult) than for the addition of two ciphertexts (CT-CT add). Our analysis also indicates that for the first multiplication to be successful, a minimum value of $L = 200$ bits is required and each additional multiplication in the circuit (increase in the multiplicative depth by 1) requires an increase in the value of $L$ by approximately 100 bits.

It must be highlighted that the impact of $L$ on the security estimate ($\lambda$) is non-linear. Note that evaluation of deeper (multiplicative depth) circuits would require large values of $L$. This in turn would require a large increase in the value of $m$ to maintain the same level of security, consequently slowing down the computations significantly as well as increasing the memory requirements (due to larger ciphertext sizes). Therefore, it is critical to constrain the depth parameter $L$ to a reasonable value and design the machine learning model to fit the depth constraint.

\subsection{Dataset and Inference Results}

We use the MNIST dataset for our experiments. This dataset consists of $60000$ $28 \times 28$ grayscale images of the $10$ digits (0-9), along with a test set of $10000$ images. We designed a simple CNN model that consists of a single convolutional layer with $28$ filters (each having a kernel size of $3 \times 3$) without any padding, followed by \emph{ActPool1} layer (polynomial approximation of ReLU and mean pooling), a flattening layer (whose output dimension is $13 \times 13 \times 28 = 4732$), a single fully connected layer of size $4732 \times 10$, and a final softmax layer. The CNN is trained using the Keras library for $10$ epochs using a {\em sparse\_categorical\_crossentropy} loss function and ADAM optimizer with a learning rate of $0.01$. The test accuracy of this simple CNN model on plaintext images was found to be $97.86$\%, which is only marginally lower than that of a CNN where the approximate ReLU function is replaced with the standard ReLU function ($98.39$\%).

Next, we focus on the problem of encrypted inference where both the inputs to the CNN and weights of the network are encrypted. To implement inference based on the above simple CNN model, the depth parameter $L$ was set to $600$ and the values of $m = 2^{16}$ and $r = 35$ were retained. The conservative security estimate based on these parameters is $128$ bits, which is obtained by using an in-built library call in HElib. Note that the inference process does not require the computation of the softmax function, which is monotonic in nature. Thus, the user can decrypt the output of the fully connected layer to infer the predicted class, which is the class with the maximum output at the FC layer. Based on the above HE parameters, the number of slots ($K$) available is more than the number of test images ($10000$). Thus, the complete inference process could be executed in the encrypted domain in one batch without any loss of test accuracy ($97.86$\%).

The time required for the execution of each layer in the CNN during encrypted inference is shown in Table \ref{execution_time}. It must be emphasized that the reported timings in Table \ref{execution_time} are for evaluating a single filter on a single thread. The results for the complete network (all $28$ filters) are shown in the next sub-section on multi-threading. As one would expect, the convolution layer is primary bottleneck in terms of number of computations. On the other hand, the fully connected (FC) layer presents a significant challenge in terms of memory bottleneck. Even for a single filter, the FC layer requires holding ($(M-P+1)(N-Q+1)(R+1)/4$) ciphertexts in memory, where $R$ is the number of classes (($R=10$ for MNIST). The number of ciphertexts gets inflated by a factor of $28$ when all the filters are considered, which makes parallelization more difficult.

\begin{table*}[t]
\centering
\begin{center}
\begin{tabular}{|c|c|}
\hline
Operation   &  Execution Time (in seconds) for Single Filter and Single Thread\\
\hline
Convolution & 487.4 \\
\hline
Approximate ReLU & 102.1 \\
\hline
Mean pooling & 16.9 \\
\hline
Fully Connected & 123.4 \\
\hline
Total (including overhead) & 812.6 \\
\hline
\end{tabular}
\end{center}
\caption{Execution time for different layers of the proposed simple CNN model.}
\label{execution_time}
\end{table*}

\subsection{Parallelization using multi-threading}

HElib supports thread level parallelism (inherited from the underlying NTL library), which can speed up the execution time. Theoretically, the design of the proposed simple CNN allows us to compute the response of each convolution filter as well as the subsequent activation, pooling, and FC layers independently. However, in practice this was not feasible due to memory constraints. Moreover, our computing environment has 112 CPUs, but the model has only 28 filters. Therefore, to make the best use of the available resources we had to implement a nested multi-threading strategy.

In the nested multi-threading strategy, we spawn $F$ (we choose $F=14$ or $F=28$) threads for the filters, and each filter thread in turn spawns $C$ convolution threads (we choose $C=1,3,5$ or $7$). Each convolutional thread operates on a horizontal partition of the full image. For example, when $C=7$, the $28 \times 28$ image is partitioned into $7$ horizontal sub-images with appropriate overlap (to avoid boundary issues). Since the approximate RELU is a pixel-wise operation, it can be applied to each horizontal partition independently. But since the mean pool operation could span across partitions, we have to wait for all the convolution threads to complete before meanpool is applied. Finally, due to memory constraints, we could not proceed with the FC layer directly. Instead, we waited for all the filter threads to complete and then spawn $H$ new threads, one for each of the $10$ classes (columns of the FC layer matrix). As earlier, these $H$ class threads further spawn $J$ channel threads to parallelize the matrix multiplication task along rows of the FC layer matrix.

\begin{figure}
    \begin{center}
    \includegraphics[width=0.5\textwidth]{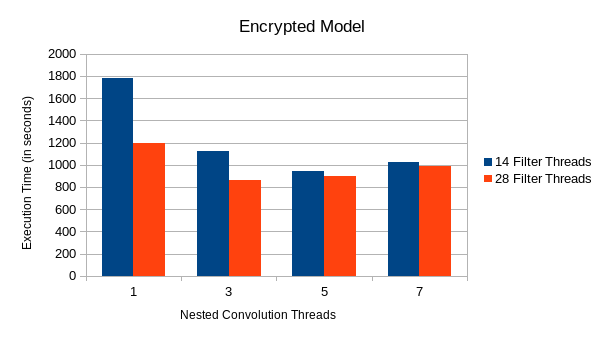}
    \caption{Execution time for various number of filter and convolution threads. In this experiment, we fix the number of class and channel threads to $10$ and $1$, respectively.}\label{fig:MultiThreadedTimings}
    \end{center}
\end{figure}

Figure \ref{fig:MultiThreadedTimings} shows the execution times for the above nested multi-threading strategy. For this experiment, we fix $H=10$ and $J=1$ and vary $F$ and $C$. From this figure, we can observe that having more filter threads generally leads to faster execution. Since the maximum number of available threads is 112, we observe that utilizing only 70-80 threads provides the optimal results because it achieves the best compromise between the number of threads and thread efficiency. In particular, when $F = 28$ and $C = 3$, 84 threads are utilized in total and it results in the lowest execution time of $864$ seconds. Similarly, when the number of channel threads $J$ is increased, the best results were obtained when $J=7$ leading to a total execution time of $561$ seconds.

Thus, the proposed multi-threading strategy leads to approximately $40$ times improvement over the time required for computing all the $28$ filters on a single thread ($28 \times 812 [Table 1] = 22736$ seconds). This is achieved by utilizing $70-80$ threads on average. If the memory constraints can be surmounted, there is scope for further refinement in the proposed multi-threaded strategy.

It must be emphasized that the values reported in the above paragraphs are the actual execution times and not the amortized time. Since $K$ images are processed in parallel exploiting the SIMD mechanism, the amortized time per image could be orders of magnitude lower. Since $K = 16384$ in our experiments, the amortized execution time for inference on a single image can be estimated as $34$ milliseconds. Note that most of the reported literature on inference based on encrypted MNIST data (e.g., \cite{gilad2016cryptonets,fastercryptonets}) use the BFV encryption scheme available in the Microsoft SEAL library implemented on a x86 platform. Hence, it is not possible to directly compare the execution times reported in this work with those reported in the literature. However, we do note that the wall-clock run time reported in \cite{gilad2016cryptonets} is 250 seconds, with only 10 $5 \times 5$ filters and a stride length of 2, which is comparable to our run time of 561 seconds (with 28 filters and a stride length of 1). Moreover, unlike \cite{gilad2016cryptonets}, the weights of the convolution and fully-connected layers are encrypted in our proposed implementation.

It must be highlighted that the ciphertext packing strategy used in this work has high latency for a single inference. This could be addressed using different packing strategies as discussed in \cite{brutzkus2019low}, which can reduce the latency at the cost of decreasing the throughput. For example, for single image inference with a different packing strategy (the entire image is encrypted within a single ciphertext), we were able to achieve a total inference time of 8.8 seconds with encrypted model parameters. When the model parameters are not encrypted, the latency can be further reduced to 2.5 seconds, which is comparable to the latency of 2.2 seconds reported in \cite{brutzkus2019low}.

\section{Conclusions and Future Work}
\label{sec:conclusion}
In this paper, we attempted to address the computation complexity in a convolutional deep neural network for encrypted processing. The approach presented in the paper describes a secure way to compute convolution, non-linear activation and pooling layers using power of the CKKS FHE scheme. We exploited two features to tame the complexity: SIMD ciphertext packing and thread level parallelism. In addition, we empirically set the depth of the circuit to further minimize the execution time. Using a recent method to approximate ReLU, we were able to build a bootstrap-free ReLU function in FHE. Since the max pooling layer requires bootstrapping, we replaced it with a mean pooling layer. Overall, we show that it is possible to implement encrypted inference in reasonable time by investing in the right CNN design and parameter choices.

In future, we will proceed to integrate other layers with the end-goal of support full inferencing and training on encrypted data. We also observe that an overall latency of $561$ seconds could be too high for some applications. The limitation of the ciphertext packing strategy used in this work is high latency for a single inference and high memory requirement. This could be addressed using different packing strategies as discussed in \cite{brutzkus2019low}, which can reduce the latency while also decreasing the throughput. Since all the ideas contained in \cite{brutzkus2019low} are practical HE implementation nuances that do not require any change to the machine learning model, they can be readily applied to the proposed approach. We leave this analysis for future work.

\section{Acknowledgement}
\label{sec:acknowledgement}
The work was initiated when authors Nandakumar/Ratha/Pankanti were part of IBM Research. We thank IBM, USA for providing us the computing facilities. Author Kumar acknowledges Infosys Foundation for the financial support to IIIT-Bangalore through the Infosys Foundation Career Development Chair Professor.

\end{document}